\documentstyle[12pt]{article}
\begin{document}                 
\title{Remarks on time-energy uncertainty relations}
\author{Romeo Brunetti and Klaus Fredenhagen \\
II Inst. f. Theoretische Physik, Universit\"at Hamburg, \\
149 Luruper Chaussee, \\
D-22761 Hamburg, Germany.}

\maketitle

\vskip20truept
Dedicated to Huzihiro Araki on the occasion of his seventieth birthday
\vskip20truept

\begin{abstract} 
Using a recent construction of observables characterizing 
the time of occurence of an effect in quantum theory, we 
present a rigorous
derivation of the standard time-energy uncertainty relation. 
In addition, we prove an uncertainty relation for time 
meaurements alone.  
\end{abstract}

\section{Introduction}\label{sec:intro}
Time-energy uncertainty relations played an important 
r\^ole in the early discussions on the physical interpretation 
of quantum theory \cite{H}. But contrary to the position-momentum 
uncertainty relation, their derivation and even precise 
formulation suffer from the 
difficulty of assigning an observable (in the sense of selfadjoint operators) 
of quantum theory with 
the measurement of time \cite{P}\cite{LM}. Meanwhile, as advocated long ago by 
Ludwig \cite{L}, it is widely accepted that the concept of 
observables should be generalized, by allowing not only 
selfadjoint operators (corresponding to projection valued 
measures) but also positive operator valued measures \cite{N}, and it 
was quickly realized that one can find such measures which 
transform covariantly under time translations and fulfil 
therefore all formal requirements for a time observable 
\cite{B}\cite{K}\cite{SV}\cite{W}. 
In a recent paper \cite{BF} we gave the first (to our knowledge)  
{\sl general construction} of such measures 
starting from an arbitary positive operator which may be 
interpreted as the {\sl effect} whose occurence time is described 
by the measure; this notion of time is closely related to the concept of 
time of arrival but in contrast to this our construction always leads 
to positive 
operator valued measures (cfr. the book in \cite{B}).
 All that is done in a strict quantum language, no classical
ideas or generalizations of quantum mechanics are involved
(we stress that the construction is valid also in quantum field theory, 
a subject which we plan to deal with in the future). 
In the present paper we show that for 
these time observables the usual time-energy uncertainty 
relation holds, and that in addition, provided the 
Hamiltonian is positive, one finds an uncertainty relation 
for time measurements alone which takes the form
\begin{displaymath}
        \Delta T \ge \frac{\mathrm{const}}{\langle H\rangle} 
\end{displaymath}
with a universal constant.
Arguments for these relations have been given by several 
authors \cite{A}\cite{B}\cite{PF}\cite{SV}\cite{Wig}, mainly 
in special situations and 
often only on a heuristic level. It is the aim of the present 
note to show that these relations can be rigorously derived for an 
arbitrary time translation covariant positive operator valued 
measure.

\section{Covariant Naimark-Stinespring 
Dilation}\label{sec:dilation}
Let ${\cal H}$ be a Hilbert space and $(U(t))$ a strongly 
continuous unitary group on ${\cal H}$ describing the time 
evolution. The occurence time of an effect is, following 
\cite{BF}, described by a covariant positive operator valued 
measure $F$, i.e. for any Borel subset $B$ of the real line
 $\mathbf R$ we have a positive 
contraction $F(B)$ on ${\cal H}$, and the operators $F(B)$ 
satisfy the conditions
\begin{displaymath}
        F(\bigcup_{n}B_{n}) = \sum_{n} F(B_{n})
\end{displaymath}
if the sets $B_{n}$ are pairwise disjoint and where the sum 
on the r.h.s. converges strongly,
\begin{displaymath}
        F({\mathbf R})=1 \ , \ F(\emptyset) =0 \ ,
\end{displaymath}
\begin{displaymath}
        U(t)F(B)U(-t) = F(B+t) \ .
\end{displaymath}
We want to construct from these data an extended Hilbert 
space ${\cal K}$ with a projection $P$ onto the subspace 
${\cal H}$, a unitary group $(V(t))$ on ${\cal K}$ which reduces 
on ${\cal H}$ to the original time translation $(U(t))$ and a 
covariant projection valued measure $E$ on ${\cal K}$ such that
\begin{displaymath}
        PE(B)P=F(B)P\ .
\end{displaymath}
Let ${\cal K}_{0}$ be the space of bounded piecewise continuous 
functions on $\mathbf{R}$ with values in ${\cal H}$. On this space we 
introduce the positive semidefinite scalar product
\begin{displaymath}
        \langle \Phi,\Psi \rangle = \int \bigl(\Phi(t), F(dt) 
        \Psi(t) \bigr) \ .
\end{displaymath}
${\cal H}$ can be isometrically embedded into ${\cal K}_{0}$ by identifying the 
elements of ${\cal H}$ with constant functions. The enlarged 
Hilbert space ${\cal K}$ is then defined as the completion of 
the quotient of ${\cal K}_{0}$ by the null space of the scalar 
product.

On ${\cal K}_{0}$ we define
\begin{displaymath}
        (E(B)\Phi)(t) = \left\{
        \begin{array}{ccc}
                \Phi(t) & , & t\in B  \\
                0 & , & t \not\in B
        \end{array}\right.
\end{displaymath}
\begin{displaymath}
   (V(t)\Phi)(s) = U(t)\Phi(s-t) \ ,            
\end{displaymath}       
\begin{displaymath}
        P\Phi = \int F(dt) \Phi(t) \ .
\end{displaymath}
$E(B)$ and $V(t)$ map the null space into itself, $P$ 
annihilates it. Therefore they are well defined operators 
on ${\cal K}$ and form the desired covariant dilation. In 
particular, $(E,V)$ is a system of imprimitivity over $\mathbf{R}$
\cite{V} 
and therefore unitarily equivalent to a multiple of the 
Schr\"odinger representation.
\section{The general Time-Energy Uncertainty 
Relation}\label{sec:time-energy}
Using the covariant dilation described in the previous 
section we can use the standard uncertainty relation in the 
Schr\"odinger representation on $L^2(\mathbf{R})$,
\begin{displaymath}
        \Delta_{\Phi}(x) \Delta_{\Phi}(\frac{1}{i}\frac{d}{dx}) \ge 
        \frac{1}{2}
\end{displaymath}
which holds for all wave functions $\Phi$ in the 
intersection of the domains of $x$ and 
$\frac{1}{i}\frac{d}{dx}$. This follows from the validity 
of the canonical commutation relations in the sense of 
quadratic forms,
\begin{displaymath}
        \bigl(x\Phi, \frac{1}{i}\frac{d}{dx}\Psi \bigr) -
        \bigl(\frac{1}{i}\frac{d}{dx}\Phi, x\Psi \bigr) = i 
        \bigl(\Phi, \Psi \bigr) 
\end{displaymath}
which may be derived from the fact that 
$\frac{1}{i}\frac{d}{dx}$ is the generator of translations 
$U(a)$, 
and that, by Stone's Theorem \cite{RS}, $a\mapsto U(a)\Phi$ is 
strongly differentiable for $\Phi$ in the domain of 
$\frac{1}{i}\frac{d}{dx}$. Namely, we have
\begin{displaymath}
\bigl(x\Phi, \frac{1}{i}\frac{d}{dx}\Psi \bigr) -
        \bigl(\frac{1}{i}\frac{d}{dx}\Phi, x\Psi \bigr)
\end{displaymath}
\begin{displaymath}
        =\frac{1}{i}\frac{d}{da}|_{a=0}\left( \bigl(x\Phi,U(a)\Psi\bigr) - 
        \bigl(U(-a)\Phi,x\Psi\bigr) \right)
\end{displaymath}
\begin{displaymath}
        =\frac{1}{i}\frac{d}{da}|_{a=0} \left( \bigl( 
        x\Phi,U(a)\Psi \bigr) - 
        \bigl(\Phi,(x+a)U(a)\Psi\bigr)\right)
\end{displaymath}
\begin{displaymath}
        =\frac{1}{i}\frac{d}{da}|_{a=0} 
        (-a)\bigl(\Phi,U(a)\Psi\bigr) = i \bigl(\Phi,\Psi\bigr) \ .
\end{displaymath}
We can now state the general time-energy uncertainty 
relation:
\begin{quote}{\it
        Let $F$ be a time translation covariant positive operator 
        valued measure, and let $H$ denote the Hamiltonian. Let 
        $\Phi$ be a unit vector in the domain of the Hamiltonian 
        for which the second moment of the probability measure 
        $d\mu(t)= \bigl( 
        \Phi,F(dt)\Phi\bigr)$ is finite. Then we have the 
        uncertainty relation
        \begin{displaymath}
                \Delta_{\Phi}(T_{F}) \Delta_{\Phi}(H) \ge \frac{1}{2}
        \end{displaymath}
        where $\Delta_{\Phi}(T_{F})$ is the square root of the 
        variance of $\mu$ and $\Delta_{\Phi}(H)=(||H\Phi||^2 
        -(\Phi,H\Phi)^2)^{\frac{1}{2}}$ is the usual energy 
        uncertainty.}
\end{quote}
\noindent{\em Proof:} We use the covariant dilation described in 
Section 2. For $\Phi\in{\cal H}$ we have 
\begin{displaymath}
        \bigl( \Phi,E(B) \Phi \bigr) = \bigl( \Phi,F(B) \Phi 
        \bigr) , 
\end{displaymath}
hence $\Phi$ is in the domain of definition of the 
selfadjoint operator $T_{E}$ defined by the projection 
valued measure $E$. Moreover, since the dilated time 
translations $V(t)$ restrict on ${\cal H}$ to the original time 
translations, $\Phi$ is also in the domain of the generator 
$K$ of $V$. But $K$ and $T_{E}$ satisfy the canonical 
commutation relation in the sense of quadratic forms, 
thus $\Phi$ fulfils the uncertainty relations with respect 
to $T_{E}$ and $K$. The desired time-energy uncertainty 
relation now simply follows from the equalities
\begin{displaymath}
        \Delta_{\Phi}(T_{E})= \Delta_{\Phi}(T_{F}) \ , \ 
        \Delta_{\Phi}(K) =      \Delta_{\Phi}(H) \ . 
\end{displaymath}

It is clear that the tricky point 
was to find a useful representation of the Hilbert space $\cal K$ with which 
we could reduce the computation to the standard position-momentum 
uncertainty relation. 
However, we stress that this representation only plays an auxiliary 
r\^ole, no physical interpretation has to be associated with it. (An
essentially equivalent derivation may already be found in \cite{W}.)

\section{Uncertainty of Time}\label{sec:time}
The replacement of projections by positive operators in the 
description of time observables leads to an intrinsic 
uncertainty. We will assume in this section that the 
Hamiltonian is positive. Under this condition, we will show 
that the minimal time 
uncertainty is inversely proportional to the expectation 
value of the energy.

Let $\Phi$ be a unit vector for which the time uncertainty 
$\Delta_{\Phi}(T_{F})$ and the expectation value of the 
Hamiltonian are finite. We use the same covariant dilation 
as before. Because of the positivity of $H$, ${\cal H}$ must 
be contained in the spectral subspace of $K$ corresponding 
to the positive real axis. We may realize ${\cal K}$ as the 
space of square integrable functions $L^2(\mathbf{R},\mathcal{L})$ 
where $\mathcal{L}$ is the Hilbert space which describes 
the multiplicity of the Schr\"odinger representation.
$K$ acts by multiplication and $T_{E}$ as generator of 
translations. Since $\Phi$ is in the domain of $T_{E}$, it 
is absolutely continuous, and since ${\cal H} \subset 
{\cal K}_{+}=L^2(\mathbf{R}_{+},\mathcal{L})$, $\Phi$ has to vanish at $x=0$. 
Hence $\Phi$ is in the quadratic form domain $q$ of the operator 
$-\frac{d^2}{dx^2}$ on ${\cal K}_{+}$ with Dirichlet boundary condition at 
$x=0$ (symbolically $-\frac{d^2}{dx^2}|_{\cal D}$). 

Since the problem is invariant under time shifts we may assume that 
the expectation value of $T_F$ vanishes, and to determine the infimum 
(over $\Phi$) of the quantity
\begin{displaymath}
        \bigl(\Phi, 
        -\frac{d^2}{dx^2}|_{\cal D}\Phi\bigr) \bigl(\Phi,x \Phi\bigr)^2 \ ,
\end{displaymath}
it would be sufficient to take it over the set
 ${\cal S}=\{\Phi\in{\cal K}_{+},||\Phi||=1, 
\Phi\in q(-\frac{d^2}{dx^2}|_{\cal D})\cap q(x)\}$. 

We use the following relation which is valid for $a,b>0$,
\begin{displaymath}
        ab^2 = \inf_{\lambda>0} \frac{4}{27\lambda^2} (a+\lambda 
        b)^3 \ .
\end{displaymath}
The relation may be verified by noting that the argument of 
the infimum assumes the value of the left side for 
$\lambda=\frac{2a}{b}$, hence it suffices to check the 
inequality 
\begin{displaymath}
        ab^2\le \frac{4}{27\lambda^2}(a+\lambda b)^3 \ , \ a,b,\lambda>0 \ .
\end{displaymath}
Setting $c=\lambda b$, we obtain the equivalent inequality
\begin{displaymath}
        a(\frac{c}{2})^2 \le (\frac{a+c}{3})^3 \ .
\end{displaymath}
Taking now the logarithm on both sides we find again an equivalent 
inequality which is a direct consequence of the concavity of 
the logarithm.

We therefore obtain the following relation
\begin{displaymath}
                \inf_{\Phi\in \mathcal{S}} \bigl(\Phi, 
        -\frac{d^2}{dx^2}|_{\cal D}\Phi\bigr) \bigl(\Phi,x 
        \Phi\bigr)^2  
\end{displaymath}
\begin{displaymath}
\qquad = \inf_{\Phi\in\mathcal{S}} \inf_{\lambda>0} 
        \frac{4}{27\lambda^2}\bigl(\Phi, 
        (-\frac{d^2}{dx^2}|_{\cal D} +\lambda x)\Phi \bigr)^3 \ .
\end{displaymath}

We may perform on the right hand side first the infimum 
over $\Phi$. We then can exploit the behaviour of the 
operator $-\frac{d^2}{dx^2}|_{\cal D} + x$ under scale transformations. 
Namely, let 
\begin{displaymath}
        (D(\mu)\Phi)(x)= \mu^{\frac{1}{2}}\Phi(\mu x)
\end{displaymath}
be the unitary scale transformations on ${\cal K}_{+}$. 
Then we have
\begin{displaymath}
        D(\lambda^{\frac{1}{3}})^{-1} (-\frac{d^2}{dx^2}|_{\cal D} + \lambda x) 
        D(\lambda^{\frac{1}{3}}) = \lambda^{\frac{2}{3}}(-\frac{d^2}{dx^2}|_{\cal D} +  x) \ .
\end{displaymath} 
Since the set $\mathcal{S}$ is scale invariant, the infimum over 
$\Phi$ is independent of $\lambda$. We thus obtain
\begin{displaymath}
                \inf_{\Phi\in \mathcal{S}} \bigl(\Phi, 
        -\frac{d^2}{dx^2}|_{\cal D}\Phi\bigr) \bigl(\Phi,x 
        \Phi\bigr)^2 =\frac{4}{27}c^3 \ ,
\end{displaymath}
where $c$ is the infimum of the spectrum of 
$-\frac{d^2}{dx^2}|_{\cal D} + x$.   

The spectrum of this operator is a pure point spectrum \cite{RS}\cite{F}. Its 
eigenfunctions are 
\begin{displaymath}
        \Phi_{n}(x)={\mathrm Ai}(x-\lambda_{n}) \ ,
\end{displaymath}
with eigenvalues $\lambda_{n}$ where $\mathrm{Ai}$ is the 
Airy function and $-\lambda_{n}$ 
are its zeros. The smallest eigenvalue is $\lambda_{1}=2.338$.
So we finally arrive at the uncertainty relation
\begin{displaymath}
        \Delta_{\Phi}(T_{F}) \ge \frac{d}{\langle H \rangle_{\Phi}}
\end{displaymath}
with $d=1.376\/$. 

Some comments are in order now:

\begin{itemize}
    \item[1.] The new relation gives a rather large bound if compared to the 
    original time-energy uncertainty, indeed we have
    \[
    \Delta_{\Phi}(T_{F})^2 
    \langle H^2\rangle_{\Phi}
    =\Delta_{\Phi}(T_{F})^2 \left(\langle H\rangle_{\Phi}^2 + 
     \Delta_{\Phi}(H)^2 \right) \ge 
    d^2+ \frac{1}{4} \ ,
    \]
    the exact largest lower bound of the left hand side being $9/4$.
    Let us also notice that the bound $d$ is universal, i.e., does not 
    depend on the details of the Hamiltonian $H$.

    \item[2.] The stated relation is covariant, i.e., energy shifts 
    do not change it. In case the infimum of the Hamiltonian is not 
    zero we may change $H$ with $H-\inf(\sigma(H))\cdot \mathbf{1}$, 
    where $\sigma(A)$ is the spectrum of the operator $A$ and 
    $\mathbf{1}$ is the unit operator on the Hilbert space.
    
    \item[3.] We have an explicit formula for the state with 
    minimum uncertainty, namely the state 
    $\Phi_{1}(x)=\mbox{Ai}(x-\lambda_{1})$. Its shape shows 
    how the energy spectrum has to be 
    distributed in order to have minimal dispersion in time.  
    (Recall that the variable $x$ labels the energy of the system.)
    
    \item[4.] In the light of the last remark one wonders whether it 
    would be possible to prepare such a kind of state in a laboratory 
    and check the relation explicitely.
    
    \item[5.] The same relation holds for the radial momentum of the 
    system in place of $T_{F}$ and by replacing the Hamiltonian by the 
    radius. 
\end{itemize}

\end{document}